\documentclass[10pt, conference, letterpaper]{ IEEEtran}
\IEEEoverridecommandlockouts

\usepackage[nospace]{cite}
\usepackage{amsmath,amssymb,amsfonts}
\usepackage{algorithm}
\usepackage{algorithmic}
\usepackage{graphicx}
\usepackage{booktabs}
\usepackage{url}
\usepackage{balance}
\usepackage{tikz}
\usetikzlibrary{arrows.meta,backgrounds,calc,fit,positioning}

\newcommand{\system}{PlaceMoE}
\definecolor{coreblue}{HTML}{4C78A8}
\definecolor{coreorange}{HTML}{F28E2B}
\definecolor{coregreen}{HTML}{59A14F}
\definecolor{coregray}{HTML}{4D5966}

\begin{document}

\title{\system{}: Accelerating MoE Training through Expert Placement and Replica Coordination}

\author{
\IEEEauthorblockN{
Zhiqiang Tan\textsuperscript{1},
Maoxin Wang\textsuperscript{1},
Wenxiang Lin\textsuperscript{1},
Sijie Wang\textsuperscript{1},
Qiang Wang\textsuperscript{1},
Shaohuai Shi\textsuperscript{1,*}, and
Bo Li\textsuperscript{2}}
\IEEEauthorblockA{
\textsuperscript{1}School of Computer Science and Technology, Harbin Institute of Technology, Shenzhen\\
\textsuperscript{2}Hong Kong University of Science and Technology, Hong Kong\\
Email: \{tanzhiqiang,25s151169,wenxianglin,sijiewang\}@stu.hit.edu.cn\\
\{qiang.wang,shaohuais\}@hit.edu.cn, bli@cse.ust.hk
}
\thanks{\textsuperscript{*}Corresponding author: Shaohuai Shi.}
}

\maketitle

\begin{abstract}
Distributed training of sparsely activated mixture-of-experts (MoE) models is
often bottlenecked by workload imbalance and all-to-all (A2A) communication
induced by dynamic token routing. Existing token deduplication, expert
placement, and replication techniques alleviate these bottlenecks but do not
jointly capture their distinct load semantics. Under hierarchical token
deduplication, communication counts each token once per destination group at
each hierarchy level, whereas expert computation processes every
token--expert assignment. With replication, both costs depend jointly on
physical layout and token-to-copy mapping, making separate optimization
suboptimal. We present \system{} to address this coupling. First, we formulate
the joint optimization of expert layout and source-aware token-to-copy mapping
for $E$ experts distributed across $P$ devices, accounting for both
hierarchical communication traffic and expert-compute load. Second, we develop
a solver that combines co-selection-aware expert grouping, bounded dynamic
programming for replica allocation, and alternating refinement of layout and
mapping. Third, \system{} overlaps replica-gradient synchronization with
attention backward. We evaluate \system{} on a 32-GPU cluster and a 64-NPU
cluster using Qwen3-VL and DeepSeek-V3. Experimental results show that
\system{} achieves up to a $6.94\times$ speedup in A2A communication and
$1.74$--$2.33\times$ speedups in end-to-end training over the state-of-the-art
training framework VeOmni.
\end{abstract}

\begin{IEEEkeywords}
mixture-of-experts training, expert replication and placement, hierarchical all-to-all communication
\end{IEEEkeywords}

\section{Introduction}
Mixture-of-experts (MoE) models scale neural-network capacity through sparse
activation: a router selects only a small set of top-$k$ experts for each token,
allowing the parameter count to grow without proportionally increasing the
computation per token
\cite{shazeer2017moe,lepikhin2021gshard,fedus2022switch}. In large-scale
training, these experts are commonly placed across GPUs using expert
parallelism (EP)~\cite{lepikhin2021gshard}. In the MoE layer of EP, the experts are activated according to the routing function (which is also a small learnable network) and the token dispatching is typically implemented through an all-to-all (A2A) operation~\cite{lepikhin2021gshard,hwang2023tutel}. After expert computation from the dispatched tokens, the MoE layer then
combines the outputs to the token sources through another A2A~\cite{rajbhandari2022deepspeedmoe,hwang2023tutel}. This mechanism, however, easily causes load imbalance in both communication and computation~\cite{hwang2023tutel, he2022fastermoe, liu2026laermoe}. That is, some experts (also called \textit{hot experts}) receive many tokens to compute, while the other experts require only several tokens according to the routing function. The load imbalance limits the scalability of distributed training of MoE models~\cite{he2022fastermoe,li2023lina,liu2023janus,deepseek2025v3,liu2026laermoe}.

Prior work addresses this imbalance through algorithmic or system
approaches. Algorithmic approaches, including GShard~\cite{lepikhin2021gshard}, Switch~\cite{fedus2022switch}, BASE~\cite{lewis2021base}, and others~\cite{zhou2022expertchoice,chen2022tamoe}, add auxiliary load-balancing objectives to router training. Because these objectives modify the next-token training loss, they may affect model generalization~\cite{jiang2025moecap,nguyen2026leastloaded}. System approaches instead reduce A2A traffic, schedule A2A communication, or overlap it with expert computation~\cite{he2022fastermoe,shi2023pipemoe,shi2024schemoe,pan2024parm,pan2025fsmoe,
wang2025spmoe,lin2025mast,pan2025smaware,li2023lina,liu2023janus,zhai2023smartmoe,jiang2024lancet,zhang2025comet}.

Among the approaches most closely related to this work, token deduplication~\cite{deepseek2025v3,lin2026hiermoe}, expert swapping or placement~\cite{zhai2023smartmoe,liu2023janus,nie2023flexmoe,skiadopoulos2026symi,lin2026hiermoe,nguyen2026leastloaded}, and expert replication~\cite{liu2026laermoe} constitute three major techniques for mitigating load imbalance in MoE training. Nevertheless, each of these techniques has important limitations. First, expert swapping requires runtime state migration and online planning, whose overhead can offset the benefit of improved load balance~\cite{lin2026hiermoe,liu2023janus}. Second, expert replication adds gradient communication because physical copies must aggregate their gradients before each optimizer update~\cite{skiadopoulos2026symi}. This synchronization can reduce the benefit of improved compute balance. Third, token deduplication gives communication and expert computation different load semantics. Optimizing either cost in isolation may therefore lead to suboptimal end-to-end performance.

In this work, we propose \system{}, a load-balanced training approach for MoE models that combines the benefits of token deduplication, expert placement, and expert replication. Existing hierarchical token-deduplication methods, such as HierMoE~\cite{lin2026hiermoe}, transfer only one copy of a token when that token is routed to multiple experts located within the same hardware hierarchy, such as a single GPU or multiple GPUs within the same node. The effectiveness of this deduplication depends on the physical expert layout, denoted by $L$. With expert replication, it also depends on the token-to-copy mapping $M$: each token must be dispatched to an appropriate physical copy while avoiding redundant communication and preserving balanced expert computation. The central problem is therefore: \textit{Given $P$ GPUs participating in expert parallelism, how can we jointly determine a physical expert layout $L$ and a source-aware token-to-copy mapping $M$ such that redundant token transfers during all-to-all communication are eliminated, each token--expert assignment is executed exactly once, and both communication and computation workloads are balanced?}

Jointly optimizing $L$ and $M$ is challenging. $L$ determines the number and locations of physical copies, whereas $M$ selects one available copy for each source rank and requested expert. Exhaustive search is combinatorial, and evaluating each pair requires recovering every token's destination groups at each hierarchy level and the non-deduplicated load on every rank. Pairwise relocation or hot-expert replication reduces the search space but cannot directly coordinate groups of frequently co-selected experts and may require a long sequence of decisions to reach a satisfactory solution. We therefore require a structured search strategy that preserves expert co-selection affinity while preventing excessive computational imbalance.

\system{} profiles token-level routes and calibrates a joint
communication-and-computation cost model. It uses expert demand and
co-selection affinity to generate replica allocations and hierarchy-aware
layouts, and then refines the source-aware token-to-copy mapping to balance
communication reuse against expert-compute load. The optimizer alternates
between $L$ and $M$ and evaluates the generated pairs on the profiled routes.
It can be rerun periodically as routing patterns evolve.
At runtime, $M$ is implemented as a lookup table and can be refreshed without
moving expert states. Updating $L$ also relocates expert and optimizer states
and is therefore performed less frequently. \system{} overlaps
replica-gradient synchronization with same-layer attention backward.

We evaluate \system{} on 32-GPU and 64-NPU clusters using Qwen3-VL~\cite{bai2025qwen3vl} and DeepSeek-V3~\cite{deepseek2025v3} with ShareGPT4V~\cite{chen2024sharegpt4v} and Tulu-3~\cite{lambert2025tulu3}. Compared with the original VeOmni system~\cite{ma2025veomni}, \system{} achieves up to a \textbf{$6.94\times$} speedup in A2A communication and \textbf{$1.74$--$2.33\times$} speedups in end-to-end training. When compared with the strongest baseline built on the same hierarchical token-deduplicated A2A runtime, \system{} still improves end-to-end training performance by \textbf{$1.05$--$1.25\times$}.

The main technical contributions of this paper are:
\begin{itemize}
\item We formulate the coupled effects of expert replication, physical expert layout, and token-to-copy mapping on hierarchical communication traffic and expert-compute load. We show that aggregate per-expert demand is insufficient once hierarchical token deduplication is applied.

\item We develop an affinity-aware optimization framework for replica allocation, hierarchical expert placement, and token-to-copy mapping. The framework groups frequently co-selected experts under capacity constraints, uses a bounded dynamic program to identify promising replica allocations, and iteratively refines the physical layout $L$ and token-to-copy mapping $M$ through polynomial-time updates.

\item We implement \system{} as an end-to-end MoE training system with profile-guided layout and mapping updates. $M$ is implemented as a lightweight lookup table for dispatch, while replica-gradient synchronization is overlapped with attention backward.

\item Experiments on a 32-GPU cluster and a 64-NPU cluster using two representative MoE models and two training datasets show that \system{} accelerates A2A communication by up to $6.94\times$ and end-to-end training by up to $2.33\times$ over VeOmni, while outperforming the strongest same-runtime baseline by up to $1.25\times$.
\end{itemize}

\section{Preliminaries and Motivation}
\begin{figure}[t]
  \centering
  \begin{minipage}[t]{0.49\linewidth}
    \centering
    \includegraphics[width=\linewidth]{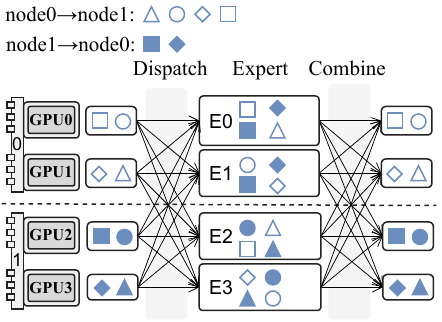}
    \par\vspace{1mm}{\footnotesize (a) Routing pattern A.}
  \end{minipage}
  \hfill
  \begin{minipage}[t]{0.49\linewidth}
    \centering
    \includegraphics[width=\linewidth]{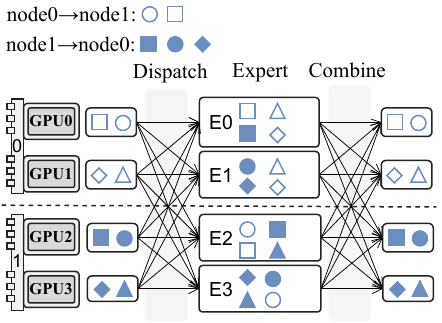}
    \par\vspace{1mm}{\footnotesize (b) Routing pattern B.}
  \end{minipage}
  \caption{Aggregate load statistics cannot characterize deduplicated
  communication. The two routing patterns have identical per-expert and
  per-rank compute loads, but different token--expert co-occurrences produce
  different inter-node traffic after deduplication.}
  \label{fig:motivation}
\end{figure}

\subsection{MoE Layers and Expert Parallelism}

\paragraph{MoE layers.}
A sparse MoE layer consists of a router and $E$ logical experts
\cite{shazeer2017moe,lepikhin2021gshard,fedus2022switch}. For each input
token $t$, the router selects a top-$k$ expert set $\mathcal{E}_t$. The
selected experts process the token independently, and their outputs are
combined using the router weights. We call each token--expert assignment an
\emph{assigned token}; hence, a routing snapshot with
$|\mathcal{T}|$ tokens contains $|\mathcal{T}|k$ assigned tokens. We preserve
the logical top-$k$ routes and optimize only where they are physically
executed.

\paragraph{Expert parallelism.}
Expert parallelism (EP) partitions experts across $P$ ranks
\cite{rajbhandari2022deepspeedmoe,hwang2023tutel}. As illustrated in
Fig.~\ref{fig:motivation}, an MoE layer uses a dispatch A2A to send tokens to
the ranks hosting their selected experts. After local expert computation, a
combine A2A returns the outputs to the source ranks for weighted aggregation.
Backpropagation performs the corresponding communication and computation in
reverse. Because these phases
are synchronous, the busiest communication group or compute rank determines
the layer's critical path.

Conventional EP places one physical copy of each logical expert. With expert
replication, a logical expert may have multiple physical copies, and each
assigned token must be mapped to one of them. This source-aware token-to-copy
mapping determines both the communication groups visited by each token and
the assigned-token load of each rank. Logical routing, physical placement,
and token-to-copy mapping are therefore distinct but coupled decisions.

\subsection{Performance Model}

\paragraph{Communication model.}
We use the standard $\alpha$--$\beta$ model for collective communication
\cite{hwang2023tutel,lin2026hiermoe}. Transferring an $m$-byte message at
communication level $h$ takes
\begin{equation}
  T_h(m)=\alpha_h+\beta_h m,
  \label{eq:alpha-beta}
\end{equation}
where $\alpha_h$ captures collective startup and synchronization latency, and
$\beta_h$ is the per-byte transfer time. Different network levels, such as the
inter-node fabric and intra-node links, use separate
$(\alpha_h,\beta_h)$ pairs~\cite{lin2026hiermoe}. Because a synchronous
collective completes only after its slowest participant, we model its critical
path using the maximum communication load rather than the average.

\paragraph{Compute model.}
Let $A_r$ denote the number of assigned tokens processed by all physical
experts on rank $r$. Unlike communication, expert computation is not
deduplicated: every token--expert assignment must be executed. We therefore
model its critical path using the busiest rank's non-deduplicated load:
\begin{equation}
  C_{\mathrm{comp}}
  =
  \gamma \max_{r\in\mathcal{R}} A_r+\delta,
  \label{eq:compute-cost}
\end{equation}
where $\gamma$ is the marginal compute time per assigned token and $\delta$
captures fixed overheads such as kernel launches. We calibrate $\alpha_h$,
$\beta_h$, $\gamma$, and $\delta$ on the target cluster
\cite{he2022fastermoe,zhai2023smartmoe,lin2026hiermoe}.

\subsection{Hierarchical Token Deduplication}

\begin{figure}[t]
  \centering
  \includegraphics[width=\columnwidth]{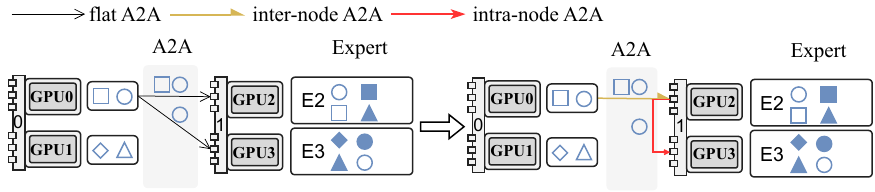}
  \caption{Hierarchical token deduplication across 2 nodes and 16 GPUs. At
  each communication level, assignments of the same token to experts in one
  destination group share a single transfer.}
  \label{fig:hierarchical-dedup}
\end{figure}

Modern accelerator clusters expose multiple communication levels. For
example, ranks within a node are connected by high-bandwidth local links,
whereas nodes communicate through a slower inter-node fabric. Let
$H$ be the number of communication levels, $\mathcal{G}_h$ the groups at level
$h$, and $\phi_h(r)$ the group containing rank $r$. A flat A2A exchanges data
directly among all $P$ EP ranks. HierMoE instead decomposes the exchange into
hierarchical stages: dispatch transfers tokens between higher-level groups
and then within each destination group, while combine follows the reverse
path \cite{lin2026hiermoe}. Figure~\ref{fig:hierarchical-dedup} also shows how
this hierarchy removes duplicate transfers. If two experts selected by token
$t$ reside in the same destination group, dispatch sends $t$ once to that
group rather than separately for each expert. The receiver expands this copy
according to the logical routing results, and combine performs the
corresponding aggregation in the reverse direction.

Given a layout $L$, let $\pi(s)$ denote the rank containing physical copy $s$.
Let $r_t$ be the source rank of token $t$, and let $M(r,e)$ select the physical
copy used when source rank $r$ requests logical expert $e$. Token $t$ visits
group $g$ at level $h$ if
\begin{equation}
  z_{t,h,g}(L,M)
  =
  \bigvee_{e\in\mathcal{E}_t}
  \mathbf{1}
  \left[
    \phi_h\!\left(\pi(M(r_t,e))\right)=g
  \right],
  \label{eq:group-incidence}
\end{equation}
and the duplicate-free token load of that group is
\begin{equation}
  U_{h,g}(L,M)=\sum_{t\in\mathcal{T}} z_{t,h,g}(L,M).
  \label{eq:deduplicated-group-load}
\end{equation}
The logical OR in Eq.~\eqref{eq:group-incidence} shows why summing per-expert
token counts cannot recover the duplicate-free communication load. For each
dispatch or combine phase $p$, we further account for its communication
direction and source group and denote the resulting exact load by
$U_{p,h,g}$.

We then apply Eq.~\eqref{eq:alpha-beta} at every communication level. Let
$\rho_{p,h}$ denote the bytes transferred per token in phase $p$ at level $h$,
as determined by the tensor width, data type, and participating ranks. The
critical-path cost of phase $p$ is
\begin{equation}
  C_{\mathrm{comm}}^{(p)}
  =
  \sum_{h=1}^{H}
  \left(
    \alpha_h
    +
    \beta_h\rho_{p,h}
    \max_{g\in\mathcal{G}_h} U_{p,h,g}
  \right).
  \label{eq:hierarchical-comm-cost}
\end{equation}
The communication cost of an MoE training iteration includes forward and
backward dispatch and combine. Let $\mathcal{P}$ denote these 4 phases; then
\begin{equation}
  C_{\mathrm{comm}}
  =
  \sum_{p\in\mathcal{P}} C_{\mathrm{comm}}^{(p)}.
  \label{eq:total-communication-cost}
\end{equation}
This model captures both heterogeneous link performance across levels and the
straggler effect of the busiest group.

\subsection{Motivations}

\begin{figure}[t]
  \centering
  \includegraphics[width=\columnwidth]{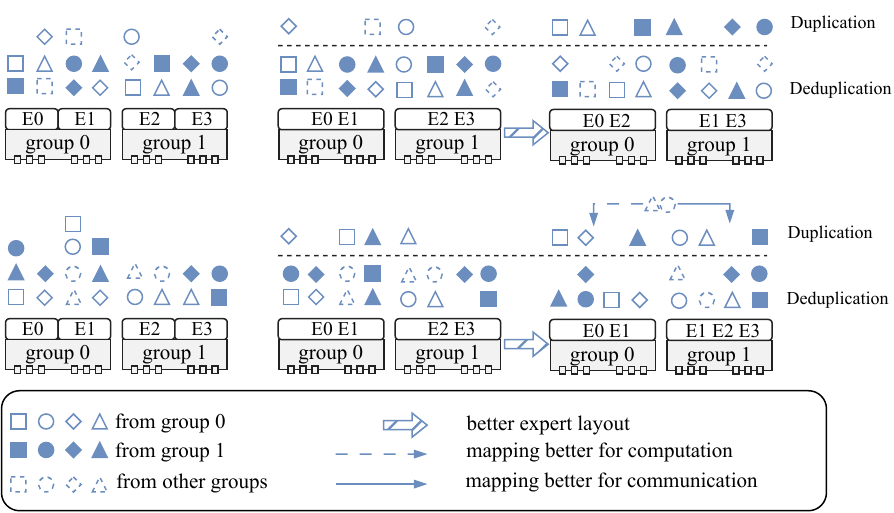}
  \caption{Why expert placement and token-to-copy mapping must be jointly
  optimized. Hollow, solid, and dashed tokens originate from group 0, group 1,
  and other groups, respectively. The upper example isolates the effect of
  placement on duplicate-free communication; the lower example shows the
  communication--compute tradeoff induced by mapping.}
  \label{fig:placement-mapping-example}
\end{figure}

Figure~\ref{fig:placement-mapping-example} illustrates why expert placement and
token-to-copy mapping must be jointly optimized. In the upper example, the two
placements have identical per-group assigned-token loads. Co-locating experts
frequently selected by the same token nevertheless reduces duplicate-free
inter-group transfers without changing compute balance.
In the lower example, $E_1$ has a copy in each group. Mapping the dashed
tokens' $E_1$ assignments to group 1 adds no new destination-group transfer
because their other selected experts already visit that group, but it
increases the compute load of group 1. Mapping them to group 0 improves
compute balance but adds another destination group. Thus, layout $L$ and
mapping $M$ jointly determine the duplicate-free communication load and
non-deduplicated expert-compute load.

Let $S$ be the number of physical slots and $n_e(L)$ the number of copies of
expert $e$ under layout $L$. For a fixed $L$, each of the $P$ source ranks
selects one of the $n_e(L)$ copies for every logical expert. The joint search
space is therefore bounded by
\[
  \sum_{L\ \mathrm{feasible}}
  \left(
    \prod_{e\in\mathcal{E}}n_e(L)
  \right)^{P}.
\]
The number of layouts grows exponentially with $S$, while the mapping choices
multiply across source ranks and experts. Moreover, evaluating each pair
requires replaying the token-level group unions in
Eq.~\eqref{eq:group-incidence}. Exhaustive joint optimization is therefore
impractical.

\section{\system{} Design}\label{sec:systemdesign}

\subsection{System Overview}

\system{} operates in three stages: profiling and calibration, layout and
mapping optimization, and training execution. It first profiles a short
training window under the default or current layout and collects a set of
token-level routing snapshots $\mathcal{D}$. Topology-aware A2A microbenchmarks
calibrate the communication model, while measured expert-compute times
calibrate the computation model. The calibrated models are reused under the
same hardware, model, and parallel configuration, while routing profiles can
be refreshed during training.

Given $\mathcal{D}$, the replica budget, and the physical topology, the
optimizer generates candidate replica allocations. For each allocation, it
constructs both item-level and group-coherent candidates, hierarchically
places physical copies across nodes and ranks, and optimizes $M$ over the
copies provided by $L$. It alternates between the two decisions for a bounded
number of rounds and selects the generated pair with the lowest calibrated
cost on the profiled routes.

At a training-step boundary, \system{} installs the selected $L$ and
implements $M$ as a lookup table for token dispatch. Updating $M$ only replaces
this table, whereas updating $L$ also relocates expert parameters and optimizer
states. Consequently, $M$ can be refreshed more frequently than $L$. During
backpropagation, \system{} aggregates gradients across physical copies and
overlaps this communication with same-layer attention backward.

\subsection{Problem Formulation}\label{subsec:formulation}

\begin{table}[t]
  \caption{Main notation}
  \label{tab:notation}
  \centering
  \footnotesize
  \begin{tabular}{@{}lp{0.72\columnwidth}@{}}
    \toprule
    Symbol & Definition \\
    \midrule
    $\mathcal{T}$ & Token set of the current MoE layer \\
    $\mathcal{E},E$ & Logical expert set and its size \\
    $\mathcal{R},P$ & EP rank set and its size \\
    $r_t,\mathcal{E}_t,\mathcal{Q}$ & Source rank and top-$k$ expert set of
      token $t$; a routing snapshot \\
    $\mathcal{D}$ & Set of profiled routing snapshots \\
    $B$ & Number of additional copies beyond one base copy per expert \\
    $d_{r,e},a_{r,e,f}$ & Profiled assignment demand and expert co-selection
      count from source rank $r$ \\
    $\mathcal{S}_r,c_r,\pi(s)$ & Physical slots on rank $r$, their capacity,
      and the rank containing slot $s$ \\
    $L(s),\mathcal{S}_e(L)$ & Expert occupying slot $s$ and the physical
      copies of expert $e$; $L(s)=\emptyset$ denotes an empty slot \\
    $M(r,e)$ & Physical copy used for tokens from source rank $r$ assigned to
      expert $e$ \\
    $H,\mathcal{G}_h,\phi_h(r)$ & Number of levels, groups at level $h$, and the group containing $r$ \\
    $U_{p,h,g},A_r$ & Duplicate-free communication load of group $g$ and
      non-deduplicated compute load of rank $r$ \\
    \bottomrule
  \end{tabular}
\end{table}

For each MoE layer, the optimizer takes the profiled routing snapshots
$\mathcal{D}=\{\mathcal{Q}^{(b)}\}_{b=1}^{|\mathcal{D}|}$,
additional-replica budget $B$, calibrated performance model, communication
hierarchy, and per-rank slot capacities as input. Each snapshot
$\mathcal{Q}=\{r_t,\mathcal{E}_t\}_{t\in\mathcal{T}}$ records every token's
source rank and logical top-$k$ experts. The optimization jointly determines a
physical expert layout $L$ and a source-aware token-to-copy mapping $M$:
\begin{equation}
\begin{aligned}
  L &: \bigcup_{r\in\mathcal{R}}\mathcal{S}_r
       \rightarrow \mathcal{E}\cup\{\emptyset\}, \\
  M &: \mathcal{R}\times\mathcal{E}\rightarrow
       \bigcup_{r\in\mathcal{R}}\mathcal{S}_r.
\end{aligned}
  \label{eq:decision-variables}
\end{equation}
Here, $\mathcal{S}_e(L)=\{s\mid L(s)=e\}$ is the set of physical copies of
expert $e$. $L$ assigns a logical expert or $\emptyset$ to every physical
slot, while $M(r,e)$ selects the copy used for tokens from source rank $r$
assigned to expert $e$. Neither decision changes $\mathcal{E}_t$ or the
logical model semantics.

Given $\mathcal{Q}$ and $(L,M)$, the preceding models yield the hierarchical
communication cost $C_{\mathrm{comm}}$ and busiest-rank compute cost
$C_{\mathrm{comp}}$. We minimize their average sum over the profiled
snapshots:
\begin{equation}
\begin{aligned}
  J_{\mathcal D}(L,M)
  ={}&
  \frac{1}{|\mathcal D|}
  \sum_{\mathcal{Q}\in\mathcal{D}}
  \Bigl[
  C_{\mathrm{comm}}(L,M;\mathcal{Q})\\[-1mm]
  &\quad+
  C_{\mathrm{comp}}(L,M;\mathcal{Q})
  \Bigr],\\
  (L^*,M^*)
  ={}&
  \arg\min_{L,M}J_{\mathcal D}(L,M).
\end{aligned}
  \label{eq:optimization-objective}
\end{equation}
Here, $C_{\mathrm{comm}}$ follows from the duplicate-free group loads in
Eq.~\eqref{eq:hierarchical-comm-cost}, while $C_{\mathrm{comp}}$ follows from
$\max_r A_r$ in Eq.~\eqref{eq:compute-cost}. Both represent synchronous
critical-path costs.

Feasibility requires $M$ to select valid copies, $L$ to respect per-rank
capacities and retain at least 1 copy of every expert, and the total number of
additional copies to equal $B$:
\begin{equation}
\begin{aligned}
  L(M(r,e)) &= e,
  &&\forall r\in\mathcal{R},e\in\mathcal{E}, \\
  \sum_{s\in\mathcal{S}_r}\mathbf{1}[L(s)\neq\emptyset]
    &\le c_r,
  &&\forall r\in\mathcal{R}, \\
  |\mathcal{S}_e(L)| &\ge 1,
  &&\forall e\in\mathcal{E}, \\
  \sum_{e\in\mathcal{E}}\bigl(|\mathcal{S}_e(L)|-1\bigr)
    &=B.
\end{aligned}
  \label{eq:feasibility-constraints}
\end{equation}

\subsection{Layout and Mapping Optimization}\label{subsec:optimization}

\system{} replaces joint enumeration with a bounded set of replica
allocations. For each allocation, capacity-constrained affinity partitioning
constructs $L$, and $M$ is optimized only over the copies in $L$. We evaluate
every retained $(L,M)$ pair on the complete profiled routes using
$J_{\mathcal D}$.

Placing frequently co-selected experts in the same communication group allows
their assignments to share a transfer. However, co-locating hot experts can
concentrate non-deduplicated assignments and create a compute bottleneck. We
therefore maximize within-group affinity while balancing assignment load
across groups, using the calibrated communication and computation costs to
quantify the tradeoff.

\textbf{Profiled demand and affinity.}
We derive demand and affinity statistics from $\mathcal{D}$. The assignment
demand from source rank $r$ to logical expert $e$ is
\begin{equation}
  d_{r,e}
  =
  \sum_{\mathcal{Q}\in\mathcal{D}}
  \sum_{t\in\mathcal{T}}
  \mathbf{1}[r_t=r,\ e\in\mathcal{E}_t],
  \label{eq:profile-demand}
\end{equation}
where $\mathcal{T}$ is the token set of snapshot $\mathcal{Q}$. The
source-conditioned affinity between experts $e$ and $f$ is
\begin{equation}
  a_{r,e,f}
  =
  \sum_{\mathcal{Q}\in\mathcal{D}}
  \sum_{t\in\mathcal{T}}
  \mathbf{1}[r_t=r,\ \{e,f\}\subseteq\mathcal{E}_t],
  \quad e\ne f.
  \label{eq:profile-affinity}
\end{equation}
$d_{r,e}$ measures non-deduplicated assignment demand, while $a_{r,e,f}$
measures how often experts $e$ and $f$ can share communication when
co-located. These statistics generate candidates but cannot recover the
complete destination-group unions of general top-$k$ routes. We therefore
evaluate final candidates on the token-level routes in $\mathcal{D}$. Reducing
the duplicate-free load by one token at level $h$ saves
$\omega_h=\sum_{p\in\mathcal{P}}\beta_h\rho_{p,h}$ time across the 4
communication phases.

\textbf{Capacity-constrained affinity partitioning.}
We use the same partitioning primitive for replica allocation and physical
placement. Each item $i$ has demand $d_i$ and pairwise affinity $a_{i,j}$. At
hierarchy level $h$, the primitive assigns every item to a group $g$ with
capacity $q_g$ spanning $p_g$ ranks by minimizing
\begin{equation}
\begin{aligned}
  J_{\mathrm{part}}^{(h)}(\mathbf y)
  ={}&
  -\omega_h\sum_{i<j}a_{i,j}\mathbf 1[y_i=y_j]
  +\gamma\max_g
     \frac{\sum_{i:y_i=g}d_i}{p_g},\\
  &\text{s.t.}\quad
  \sum_i\mathbf 1[y_i=g]=q_g,\quad\forall g .
\end{aligned}
  \label{eq:partition-objective}
\end{equation}
The first term rewards communication reuse within a group, while the second
penalizes the highest per-rank assignment load across groups. The calibrated
$\omega_h$ and $\gamma$ express both terms in time units; $\alpha_h$ and
$\delta$ are omitted because they are constant for a fixed group structure.
We use $J_{\mathrm{part}}^{(h)}$ only to generate candidates;
$J_{\mathcal D}$ selects the final result using the complete profiled routes.

Algorithm~\ref{alg:affinity-partitioning} first applies K-means to the
normalized spectral embedding of the affinity matrix
\cite{ng2001spectral}. To enforce capacities, group $g$ creates $q_g$
positions at its centroid, and linear assignment~\cite{kuhn1955hungarian}
maps items to these positions by squared embedding distance. We alternate
assignment and centroid updates until convergence or the iteration limit, and
then greedily exchange cross-group item pairs that reduce
$J_{\mathrm{part}}^{(h)}$ without changing group sizes. Multiple K-means
initializations generate alternative candidates.

\begin{algorithm}[t]
  \caption{Capacity-constrained affinity partitioning}
  \label{alg:affinity-partitioning}
  \begin{algorithmic}[1]
    \REQUIRE $\{d_i\}$, $\{a_{i,j}\}$, hierarchy level $h$,
      capacities $\{q_g\}$, normalizers $\{p_g\}$,
      calibrated $\omega_h$ and $\gamma$
    \ENSURE capacity-feasible group labels $\mathbf y$
    \STATE $\mathbf z\gets\textsc{SpectralEmbedding}(\{a_{i,j}\})$
    \STATE $\mathbf c\gets\textsc{KMeans}(\mathbf z)$
    \REPEAT
      \STATE create $q_g$ positions at centroid $\mathbf c_g$
      \STATE $\mathbf y\gets\textsc{LinearAssignment}(\mathbf z,\mathbf c)$
      \STATE recompute $\mathbf c$ from $\mathbf y$
    \UNTIL{$\mathbf y$ is unchanged or the iteration limit is reached}
    \REPEAT
      \STATE find the cross-group pair with the largest cost reduction
      \STATE exchange the pair if the reduction is positive
    \UNTIL{the exchange limit is reached or no positive reduction remains}
    \RETURN $\mathbf y$
  \end{algorithmic}
\end{algorithm}

\textbf{Selecting copy counts.}
Let $B=uE+b$, where $0\le b<E$, be the additional-replica budget. Every expert
has one mandatory base copy, while $B$ counts only additional copies. The term
\emph{base} is only an accounting convention: all copies participate
identically in placement, mapping, and gradient synchronization. We require
$B\leq\sum_{r\in\mathcal R}c_r-E$. Every expert first receives $u$ additional
copies. If $b=0$, all experts therefore have $u+1$ physical copies.
Otherwise, let $q=\gcd(E,b)$. We invoke
Algorithm~\ref{alg:affinity-partitioning} with the logical experts as input
items, demands
\[
  d_e=\sum_r d_{r,e},
  \qquad
  a_{e,f}=\sum_r a_{r,e,f},
\]
and $E/q$ groups, each with capacity $q_g=q$ and normalizer $p_g=1$.
Because this stage identifies additional copies that can avoid the most expensive
remote transfers, it uses $\omega_h$ for the coarsest communication level.
Each partitioning restart produces an expert grouping, and selecting $b/q$
groups assigns the remaining $b$ copies to exactly $b$ experts.

The affinity $a_{e,f}$ is used to form expert groups but not to rank them,
because its actual communication benefit depends on the subsequent placement
and mapping. Let the groups be $G_1,\ldots,G_m$, where $m=E/q$. A selected set
$C$ has shortlist score $\sum_{G\in C}\sum_{e\in G}d_e$. Because demand alone
cannot predict deduplicated communication, retaining only the highest-scoring
allocation may discard a better joint-cost candidate. We therefore use a
bounded dynamic program to preserve several alternatives. Let
$K_{\mathrm{alloc}}$ be the shortlist limit and
let $\mathrm{DP}[j,v]$ store the highest-scoring candidates that select $v$ of
the first $j$ groups. Starting from $\mathrm{DP}[0,0]=\{\emptyset\}$, with
$\mathrm{DP}[0,v]=\emptyset$ for $v>0$, the update is
\begin{equation}
\begin{aligned}
\mathrm{DP}[j,v]=\operatorname{Top}_{K_{\mathrm{alloc}}}\Big(&
  \mathrm{DP}[j-1,v]\ \cup\\[-1mm]
 &\{C\cup\{G_j\}:C\in\mathrm{DP}[j-1,v-1]\}\Big).
\end{aligned}
  \label{eq:replica-allocation-dp}
\end{equation}
Here, $\operatorname{Top}_{K_{\mathrm{alloc}}}$ retains at most
$K_{\mathrm{alloc}}$ candidates with the highest shortlist scores.
$\mathrm{DP}[m,b/q]$ contains candidates that assign one additional copy to
exactly $b$ experts. We merge and deduplicate the candidates from all
partitioning restarts before constructing and evaluating their $L$ and $M$.

\textbf{Initializing \(L\) and \(M\).}
For a retained allocation, let $n_e$ be the number of physical copies of
expert $e$, including its base copy. Thus, $n_e=u+2$ for experts in the
selected groups and $n_e=u+1$ otherwise. The solver creates the allocated
copies and represents unused slots as empty items. Because no mapping exists
initially, it distributes each expert's demand and affinity uniformly across
its $n_e$ copies. For a copy $s$ of expert $e$ and copy $s'$ of expert $f$,
\begin{equation}
  d_{r,s}^{(0)}=\frac{d_{r,e}}{n_e},
  \qquad
  a_{s,s'}^{(0)}
  =\frac{\sum_r a_{r,e,f}}{n_en_f}.
  \label{eq:uniform-copy-statistics}
\end{equation}
Empty items and copy pairs of the same logical expert have zero affinity. We
apply Algorithm~\ref{alg:affinity-partitioning} first across nodes using
inter-node $\omega_h$, with node slot and rank counts as $q_g$ and $p_g$, and
then within each node using intra-node $\omega_h$, $q_g=c_r$, and $p_g=1$.
Maximum-weight matching maps abstract groups to capacity-compatible physical
groups using source-local demand. A demand-ordered greedy repair enforces at
most 1 copy of each expert per rank while preserving rank capacities; infeasible
candidates are discarded. Deterministic slot assignment then produces $L_0$.

Given $L_0$, we initialize $M_0$ by processing $(r,e)$ entries in descending
$d_{r,e}$. We prefer node-local copies when available and otherwise consider
all copies of $e$, choosing the smallest projected destination-rank load and
breaking ties by rank and slot ID.

We partition logical experts into one balanced affinity group per node. For
allocations preserving uniform groupwise copy counts, bounded copy-round
permutations and maximum-weight matching propose node footprints; other
allocations use the item-level path. Block updates choose one destination per
source-node--group pair using exact route unions and calibrated projected
compute load. Rank placement and entrywise $M$ refinement follow; exact
$J_{\mathcal D}$ selects the final pair, and group node footprints persist
across later rounds.

\textbf{Given \(M\), update \(L\).}
Given $M$, we aggregate the profiled statistics onto the physical copies
selected by $M$. For nonempty copies $s$ and $s'$ of different logical
experts, we compute the source-conditioned demand, aggregate demand, and
pairwise affinity as
\begin{equation}
\begin{aligned}
  d_{r,s}^{M}
  ={}&d_{r,L(s)}\mathbf{1}[M(r,L(s))=s],\\
  d_s
  ={}&\sum_{r\in\mathcal R}d_{r,s}^{M},\\
  a_{s,s'}
  ={}&
  \sum_{r\in\mathcal R}
  a_{r,L(s),L(s')}\\
  &{}\cdot
  \mathbf{1}[M(r,L(s))=s]
  \mathbf{1}[M(r,L(s'))=s'] .
\end{aligned}
\end{equation}
Here, $d_s$ is the assignment demand served by copy $s$, and $a_{s,s'}$
counts profiled tokens mapped to both copies. Using $d_s$ and $a_{s,s'}$, we
apply the same 2-level partitioning with the corresponding $\omega_h$ at
each level to obtain the updated $L$; empty slots have zero demand and
affinity. If $y_s=g'$ assigns copy $s$ to abstract group $g'$, its matching
weight to physical group $g$ is
\begin{equation}
  w_{g',g}^{M}
  =
  \sum_{s:y_s=g'}
  \sum_{r:\phi_h(r)=g}d_{r,s}^{M}.
  \label{eq:physical-group-matching}
\end{equation}
Maximum-weight bipartite matching uses these weights, followed by the same
rank-level repair used during initialization. During a layout update, each
physical copy is treated as a movable item. After assigning the items to new
physical slots, we rewrite every entry of $M$ to follow the relocated copy
before applying Eq.~\eqref{eq:mapping-update}.

\textbf{Given \(L\), update \(M\).}
With $L$ fixed, each entry $M(r,e)$ chooses from
$\mathcal S_e(L)$. Let the profiled load induced by the current mapping be
\begin{equation}
  A_{r'}(M)
  =
  \sum_{r\in\mathcal R}\sum_{e\in\mathcal E}
  d_{r,e}\mathbf 1[\pi(M(r,e))=r'].
  \label{eq:mapping-rank-load}
\end{equation}
Changing only $M(r,e)$ to copy $s$ produces peak load
\begin{equation}
\begin{aligned}
  A_{\max}^{\,r,e\rightarrow s}(M)
  =
  \max_{r'\in\mathcal R}
  \big[
  &A_{r'}(M)
  +d_{r,e}\mathbf 1[\pi(s)=r']\\
  &-d_{r,e}\mathbf 1[\pi(M(r,e))=r']
  \big].
\end{aligned}
  \label{eq:projected-rank-load}
\end{equation}
For brevity, let $g_h(s)=\phi_h(\pi(s))$. The communication reward of choosing
$s$ is
\begin{equation}
\begin{aligned}
 F^{\mathrm{comm}}_{r,e}(s)
 ={}&
 \sum_{h=1}^{H}\omega_h
 \Bigg(
   \sum_{\substack{f\in\mathcal E\\f\ne e}}
   a_{r,e,f}
   \mathbf 1[g_h(s)=g_h(M(r,f))]\\
 &\hspace{19mm}
   +d_{r,e}\mathbf 1[g_h(s)=\phi_h(r)]
 \Bigg).
\end{aligned}
  \label{eq:mapping-communication-score}
\end{equation}
The first term rewards sharing destination groups with co-selected experts,
while the second rewards source locality. We subtract the projected peak
compute time using the calibrated coefficient $\gamma$ and update
\begin{equation}
  M(r,e)
  \leftarrow
  \arg\max_{s\in\mathcal S_e(L)}
  \left[
    F^{\mathrm{comm}}_{r,e}(s)
    -\gamma A_{\max}^{\,r,e\rightarrow s}(M)
  \right].
  \label{eq:mapping-update}
\end{equation}
We visit entries in descending demand order and commit each update
immediately, so subsequent entries observe the latest mapping and rank loads.
A mapping update stops when a full sweep changes no entry or the preset sweep
limit is reached.

\textbf{Alternation and complexity.}
For each retained replica allocation, \system{} alternates $L$ and $M$ for a
bounded number of rounds. The first round uses uniform copy statistics, while
each later round uses the statistics induced by the preceding mapping. The
solver evaluates every generated pair on $\mathcal D$ using
Eq.~\eqref{eq:optimization-objective} and retains the one with the lowest
$J_{\mathcal D}$. Algorithm~\ref{alg:layout-mapping} summarizes this
procedure.

\begin{algorithm}[t]
  \caption{Joint optimization of \(L\) and \(M\)}
  \label{alg:layout-mapping}
  \begin{algorithmic}[1]
    \REQUIRE Profiled routes $\mathcal D$, replica budget, topology,
      and calibrated cost model
    \ENSURE Layout $L$ and mapping $M$
    \STATE Derive $d$, $a$, and the replica-allocation shortlist
    \FOR{each retained replica allocation}
      \STATE Initialize copy statistics by uniform allocation
      \FOR{each layout--mapping round}
        \STATE Generate or refine item-level and group-coherent candidates
        \STATE Refine each candidate mapping using
          Eq.~\eqref{eq:mapping-update}
        \STATE Evaluate and record each pair using $J_{\mathcal D}$
        \STATE Rebuild copy statistics from the lowest-cost pair
      \ENDFOR
      \STATE Record the lowest-cost pair from all rounds
    \ENDFOR
    \RETURN the recorded pair minimizing $J_{\mathcal D}$
  \end{algorithmic}
\end{algorithm}

For $N$ input items, Algorithm~\ref{alg:affinity-partitioning} takes
$O(N^3)$ time because both dense spectral decomposition and linear assignment
are cubic. The bounded assignment and exchange rounds do not change this
order. Logical-expert partitioning therefore takes $O(E^3)$ per restart, while
a layout update takes $O(S^3)$.

With $L$ fixed, let $R_e=|\mathcal S_e(L)|$. For each source rank, a mapping
sweep evaluates $\sum_e R_e\le S$ candidate copies. Without caching,
evaluating one candidate scans $E$ experts across $H$ levels, giving a
conservative $O(PEHS)$ bound per sweep. We cache source-conditioned affinity
sums by destination group and maintain rank loads incrementally, reducing the
practical cost without changing this bound. These polynomial steps replace
the exponential joint search above. Updating $M$ requires neither spectral
decomposition nor linear assignment and is therefore cheaper than rebuilding
$L$.

\subsection{Training Execution and Replica-Gradient Synchronization Overlap}

After installing $L$ and $M$, forward dispatch directly looks up $M(r,e)$ to
select the physical copy for tokens from source rank $r$ assigned to expert
$e$.

\textbf{Hot layout and mapping updates.}
At a configurable interval, \system{} profiles a token-level routing snapshot
and asynchronously launches the $L,M$ optimizer in a separate CPU process.
Training continues with the current pair while the optimizer derives the next
one. A mapping-only update takes effect by replacing the lookup table at a
training-step boundary. For a layout update, the runtime compares the old and
new layouts across all MoE layers and groups the relocated expert parameters
and optimizer states by source--destination rank pair. Each group is packed
into contiguous buffers and transferred through batched P2P communication.
Once all destination states are ready, the runtime atomically switches $L$ and
$M$. Thus, training continues without restart, and dispatch never observes a
mapping before its expert states are available.

\textbf{Replica-gradient synchronization overlap.}

\begin{figure}[t]
  \centering
  \includegraphics[width=\columnwidth]
    {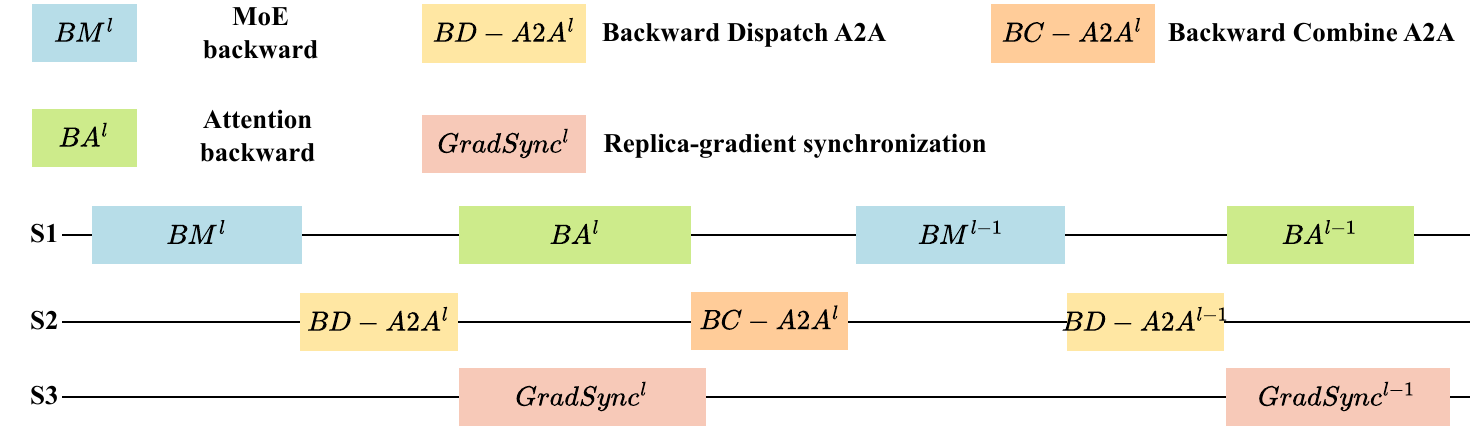}
  \caption{Replica-gradient synchronization overlap.
  Streams S1, S2, and S3 execute backward computation, MoE A2A, and
  replica-gradient communication, respectively. After
  \(BD\text{-}A2A^\ell\), \(\mathrm{GradSync}^\ell\) first overlaps
  with \(BA^\ell\).}
  \label{fig:gradient-overlap}
\end{figure}

Physical copies of the same logical expert share identical parameters but
process different assigned tokens and therefore produce different local
gradients. Prior MoE systems overlap token communication with model
computation \cite{shi2023pipemoe,pan2024parm,pan2025fsmoe,lin2025mast,
pan2025smaware}; \system{} extends this approach to replica-gradient
communication. It aggregates the gradients of all copies and writes the result
back to each copy.
Figure~\ref{fig:gradient-overlap} illustrates this overlap. We use
$BM^\ell$ and $BA^\ell$ to denote the MoE and attention backward computations
of layer $\ell$, and $BC\text{-}A2A^\ell$ and $BD\text{-}A2A^\ell$ to denote
its backward combine and dispatch A2A, respectively. Once the expert
gradients are ready and $BD\text{-}A2A^\ell$ completes, \system{} launches
$\mathrm{GradSync}^\ell$ on a separate stream, where it first overlaps with
$BA^\ell$. In Fig.~\ref{fig:gradient-overlap}, synchronization completes
within the attention-backward window. Otherwise, it continues alongside
subsequent backward work but must finish before $BD\text{-}A2A^{\ell-1}$,
after which the pipeline repeats for layer $\ell-1$. After backpropagation,
\system{} waits for any remaining synchronization tail before global
gradient-norm computation, gradient clipping, and the optimizer step.

\section{Evaluation}

\subsection{Experimental Setup}

\textbf{Testbeds.}
Table~\ref{tab:testbeds} summarizes our GPU and NPU testbeds, which use NCCL
and HCCL as their respective collective communication backends.

\begin{table}[t]
  \centering
  \caption{Testbed configurations.}
  \label{tab:testbeds}
  \footnotesize
  \begin{tabular}{@{}lp{0.63\columnwidth}@{}}
    \toprule
    Configuration & Specification \\
    \midrule
    \multicolumn{2}{@{}l}{\emph{GPU testbed}} \\
    CPU per node
      & Dual Intel Xeon Platinum 8358 @ 2.60\,GHz \\
    Accelerators
      & 8$\times$ NVIDIA RTX A6000 (48\,GB) per node \\
    Host memory
      & 512\,GB DDR4 \\
    Intra-node
      & NVLink 112.5\,GB/s (4$\times$), PCIe 4.0$\times$16 \\
    Inter-node
      & Mellanox MT28908, 200\,Gb/s \\
    Scale
      & 4 nodes, 32 GPUs \\
    Software
      & PyTorch 2.9.1+cu128, CUDA 12.8, NCCL 2.27.5 \\
    \addlinespace
    \multicolumn{2}{@{}l}{\emph{NPU testbed}} \\
    CPU per node
      & 4$\times$ Kunpeng 920 5250, 192 cores \\
    Accelerators
      & 8$\times$ Ascend 910B (64\,GB) per node \\
    Host memory
      & 1.5\,TB \\
    Intra-node
      & HCCS 392\,GB/s bidirectional \\
    Inter-node
      & 8$\times$200\,Gb/s RoCE per node \\
    Scale
      & 4/8 nodes, 32/64 NPUs \\
    Software
      & PyTorch 2.9.0, torch-npu 2.9.0.post2, CANN 9.0.0, HCCL 9.0.0 \\
    \bottomrule
  \end{tabular}
\end{table}

\textbf{Models and workloads.}
We evaluate Qwen3-VL-30B-A3B, which has 48 MoE layers, 128 experts
per layer, and top-8 routing, on ShareGPT4V for multimodal training and
Tulu-3 for text-only training. Following HierMoE's scaled-model
setup~\cite{lin2026hiermoe}, we retain 6 DeepSeek-V3 MoE layers, each with 256
experts and top-8 routing, and train them on Tulu-3.

\textbf{Baselines.}
We compare \system{} with 4 baselines.
\emph{VeOmni v0.1.11} is the state-of-the-art open-source training framework
\cite{ma2025veomni} and uses conventional A2A for token dispatch and
combination.
\emph{Replication} maintains 2 fixed copies of each logical expert, arranged
as 2 EP16 groups on 32 devices and 2 EP32 groups on 64 devices. Copies follow
a default expert order.
\emph{HierMoE} applies hierarchical token deduplication and pairwise expert
swaps~\cite{lin2026hiermoe}.
\emph{EPLB} places experts according to non-deduplicated per-expert
assignment load and represents the expert-parallel load balancer used by
DeepSeek-V3~\cite{deepseek2025v3}.
\emph{\system{}} jointly optimizes replica allocation, hierarchical expert
placement, and token-to-copy mapping.
Replication, EPLB, HierMoE, and \system{} share the same hierarchical
token-deduplicated A2A runtime, providing a common communication substrate;
VeOmni retains its original A2A path. Unless otherwise stated, Replication,
EPLB, and \system{} use the same 1-fold redundancy budget $B=E$, adding one
copy per logical expert on average. VeOmni and HierMoE use no additional
copies ($B=0$).

\textbf{Methodology.}
We set $K_{\mathrm{alloc}}=64$, use 3 partition restarts and 3
layout--mapping rounds, and limit capacity assignment to 12 iterations. Pair
refinement allows at most 24 exchanges, reduced to 12 for rank-level
partitions, while each mapping update performs at most 6 sweeps. All
procedures terminate early upon convergence. We discard the first 10 warm-up steps
and report the mean over the next 10 steady-state steps. Unless otherwise
stated, EP32 and EP64 runs use global batch sizes of 128 and 256, respectively,
and each rank processes approximately 16K input tokens. All methods use
the same model, data, numerical precision, and optimizer configuration.
End-to-end time is the mean wall-clock time per steady-state training step;
initial profiling, calibration, and layout--mapping optimization are excluded.
A2A time sums the forward and backward critical-rank A2A spans;
expert-compute time is measured analogously. All speedups are relative to
VeOmni on the same platform and workload.

\subsection{Validation of the Cost Model}

Our cost model sums the duplicate-free hierarchical A2A cost in
Eq.~\eqref{eq:hierarchical-comm-cost} and the non-deduplicated expert-compute
cost in Eq.~\eqref{eq:compute-cost}. We fit $(\alpha_h,\beta_h)$ by linear
regression over topology-aware collective microbenchmarks with different
message sizes, and fit $(\gamma,\delta)$ using the first 3 profiled training
steps.

\begin{figure}[t]
  \centering
  \includegraphics[width=\columnwidth]
    {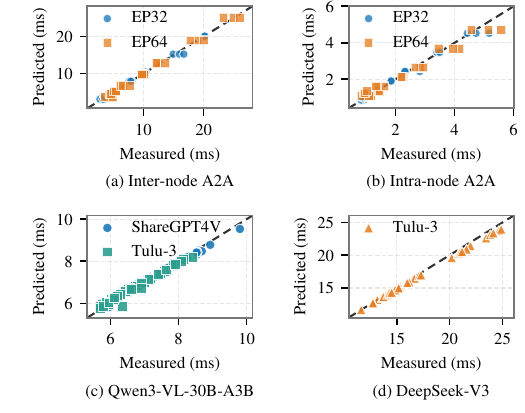}
  \caption{Predicted versus measured A2A and expert-compute times on NPUs. The
  first row validates the inter- and intra-node communication models; the
  second validates the expert-compute models. Dashed lines show ideal
  prediction.}
  \label{fig:cost-model-validation}
\end{figure}

Figure~\ref{fig:cost-model-validation} compares the predictions with held-out
NPU measurements. The inter- and intra-node models achieve $R^2$ values of
$0.9922$ and $0.9694$ and MAPEs of $4.80\%$ and $8.34\%$, respectively. For
Qwen3-VL-30B-A3B, the compute model achieves $R^2=0.9851$ and $0.9671$ on
ShareGPT4V and Tulu-3, with MAPEs of $1.38\%$ and $1.41\%$, respectively.
The Tulu-3 prediction reuses coefficients calibrated on ShareGPT4V,
demonstrating cross-dataset reuse. For DeepSeek-V3, the model achieves
$R^2=0.9869$ and a MAPE of $1.75\%$. These results validate the assumed
linear relationships. Model fitting takes less than 2 seconds and requires
only a few profiled training steps. Communication coefficients are calibrated
per cluster topology and EP configuration, while compute coefficients are
calibrated per model and execution configuration.

\subsection{End-to-End Training Performance}

\begin{figure}[t]
  \centering
  \includegraphics[width=0.98\columnwidth]
    {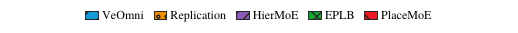}\\[-0.2em]
  \begin{minipage}[t]{0.49\columnwidth}
    \centering
    \includegraphics[width=\linewidth]
      {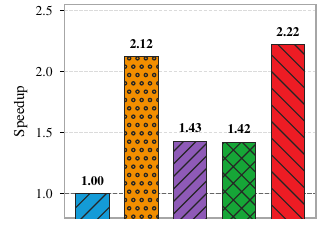}\\[-0.25em]
    \footnotesize (a) 32 GPUs, Qwen3-VL\\[-0.15em]
    \scriptsize ShareGPT4V, 16K tokens/rank
  \end{minipage}\hfill
  \begin{minipage}[t]{0.49\columnwidth}
    \centering
    \includegraphics[width=\linewidth]
      {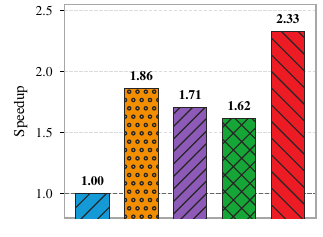}\\[-0.25em]
    \footnotesize (b) 32 NPUs, Qwen3-VL\\[-0.15em]
    \scriptsize ShareGPT4V, 16K tokens/rank
  \end{minipage}\\[0.8em]
  \begin{minipage}[t]{0.49\columnwidth}
    \centering
    \includegraphics[width=\linewidth]
      {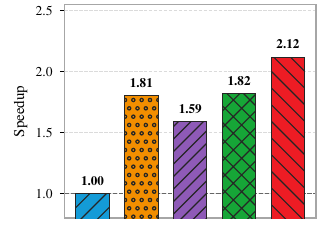}\\[-0.25em]
    \footnotesize (c) 64 NPUs, Qwen3-VL\\[-0.15em]
    \scriptsize Tulu-3, 16K tokens/rank
  \end{minipage}\hfill
  \begin{minipage}[t]{0.49\columnwidth}
    \centering
    \includegraphics[width=\linewidth]
      {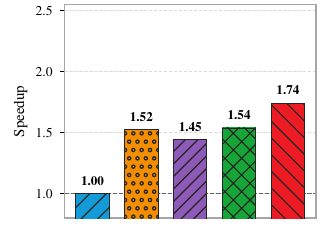}\\[-0.25em]
    \footnotesize (d) 64 NPUs, DeepSeek-V3\\[-0.15em]
    \scriptsize Tulu-3, 32K tokens/rank
  \end{minipage}
  \caption{End-to-end training speedup normalized to VeOmni; higher is
  better.}
  \label{fig:e2e-training-speedup}
\end{figure}

We evaluate \system{} across GPU and NPU clusters, 32- and 64-device scales,
text and multimodal workloads, 2 MoE models, and different token volumes.
Figure~\ref{fig:e2e-training-speedup} reports 4 representative configurations.
Across the 4 configurations, \system{} achieves $1.74$--$2.33\times$
end-to-end speedup over VeOmni. It also outperforms the strongest competing
baseline in every configuration, reducing training time by a further
$4.5\%$--$20.1\%$. Although the strongest baseline varies across workloads,
\system{} consistently performs best by jointly optimizing physical expert
layout and token-to-copy mapping under the two distinct load semantics.

\subsection{Communication and Expert-Compute Performance}

\begin{table}[t]
  \centering
  \caption{Breakdown of the 4 configurations in
  Fig.~\ref{fig:e2e-training-speedup}. Values are speedups over VeOmni;
  bold denotes the best result.}
  \label{tab:component-breakdown}
  \scriptsize
  \setlength{\tabcolsep}{2.5pt}
  \begin{tabular}{@{}llcccc@{}}
    \toprule
    Case & Metric & Replication & HierMoE & EPLB & \system{} \\
    \midrule
      & Hierarchical All-to-All
                       & 2.85 & 1.57 & 1.74 & \textbf{3.10} \\
    Fig.~\ref{fig:e2e-training-speedup}(a)
      & Expert computation & 1.05 & 0.85 & \textbf{1.11} & 1.05 \\
      & End-to-End     & 2.12 & 1.43 & 1.42 & \textbf{2.22} \\
    \addlinespace[0.2em]
      & Hierarchical All-to-All
                       & 4.19 & 3.24 & 4.06 & \textbf{6.94} \\
    Fig.~\ref{fig:e2e-training-speedup}(b)
      & Expert computation & 1.13 & 0.84 & \textbf{1.22} & 1.15 \\
      & End-to-End     & 1.86 & 1.71 & 1.62 & \textbf{2.33} \\
    \addlinespace[0.2em]
      & Hierarchical All-to-All
                       & 2.81 & 2.34 & 3.52 & \textbf{3.85} \\
    Fig.~\ref{fig:e2e-training-speedup}(c)
      & Expert computation & 1.16 & 0.95 & \textbf{1.31} & 1.26 \\
      & End-to-End     & 1.81 & 1.59 & 1.82 & \textbf{2.12} \\
    \addlinespace[0.2em]
      & Hierarchical All-to-All
                       & 2.37 & 2.09 & 3.13 & \textbf{3.43} \\
    Fig.~\ref{fig:e2e-training-speedup}(d)
      & Expert computation & 1.17 & 0.93 & 1.40 & \textbf{1.45} \\
      & End-to-End     & 1.52 & 1.45 & 1.54 & \textbf{1.74} \\
    \bottomrule
  \end{tabular}
\end{table}

Table~\ref{tab:component-breakdown} decomposes the 4 configurations in
Fig.~\ref{fig:e2e-training-speedup}. Relative to VeOmni, \system{} accelerates
hierarchical A2A by $3.10$--$6.94\times$ and expert computation by
$1.05$--$1.45\times$. It further outperforms the strongest A2A baseline in
each configuration by $1.09$--$1.66\times$. EPLB achieves
$1.11$--$1.40\times$ expert-compute and $1.74$--$4.06\times$ A2A speedups
over VeOmni. Relative to EPLB, \system{} improves A2A performance by
$1.09$--$1.78\times$ while achieving $94\%$--$104\%$ of its expert-compute
performance.

These component metrics exclude replica-gradient synchronization and therefore
do not fully determine end-to-end performance. HierMoE has no additional
copies, while Replication synchronizes regular copy pairs. EPLB's load-driven
layout creates an irregular synchronization graph and leaves this
communication on the critical path. \system{} instead overlaps
replica-gradient synchronization with same-layer attention backward.
Consequently, EPLB can improve both reported components without achieving a
comparable end-to-end gain. Figures~\ref{fig:ablation-sensitivity}(a) and (b)
show that the overlap reduces step time by $6.7\%$ on average.

These results expose the conflicting placement preferences of the two costs.
Minimizing hierarchical token-deduplicated communication favors colocating
frequently co-selected experts to share transfers, whereas balancing
non-deduplicated computation favors spreading high-demand experts across
ranks. Optimizing either objective alone is therefore insufficient: HierMoE
reduces communication but can slow expert computation, while EPLB balances
computation without exploiting token-level deduplication in placement.
\system{} improves both components over VeOmni in every configuration and
consistently achieves the best end-to-end performance, validating their joint
optimization.

\subsection{Ablation, Sensitivity, and Overhead}

\begin{figure}[t]
  \centering
  \includegraphics[width=0.98\columnwidth]
    {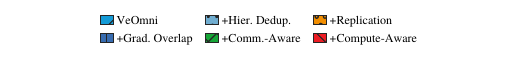}\\[-0.2em]
  \begin{minipage}[t]{0.49\columnwidth}
    \centering
    \includegraphics[width=\linewidth]
      {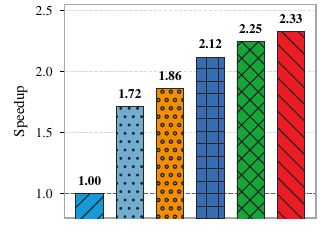}\\[-0.25em]
    \footnotesize (a) Qwen3-VL\\[-0.15em]
    \scriptsize ShareGPT4V, 16K tokens/rank
  \end{minipage}\hfill
  \begin{minipage}[t]{0.49\columnwidth}
    \centering
    \includegraphics[width=\linewidth]
      {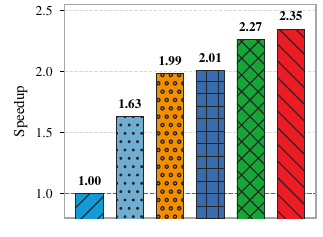}\\[-0.25em]
    \footnotesize (b) Qwen3-VL\\[-0.15em]
    \scriptsize Tulu-3, 16K tokens/rank
  \end{minipage}\\[0.8em]
  \begin{minipage}[t]{0.54\columnwidth}
    \centering
    \raisebox{0.004em}{\includegraphics[width=\linewidth]
      {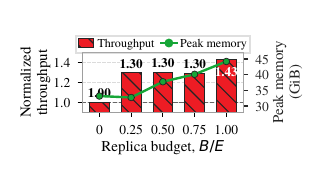}}\\[-0.25em]
    \footnotesize (c) Qwen3-VL\\[-0.15em]
    \scriptsize ShareGPT4V, 16K tokens/rank
  \end{minipage}\hfill
  \begin{minipage}[t]{0.46\columnwidth}
    \centering
    \raisebox{0.8em}{\includegraphics[width=\linewidth]
      {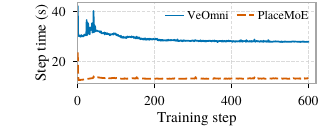}}\\[-0.25em]
    \footnotesize (d) Qwen3-VL\\[-0.15em]
    \scriptsize ShareGPT4V, 600-step run
  \end{minipage}
  \caption{Component ablations normalized to VeOmni (a--b),
  replica-budget sensitivity (c), and long-run step-time stability (d) on
  32 NPUs.}
  \label{fig:ablation-sensitivity}
\end{figure}

Figures~\ref{fig:ablation-sensitivity}(a) and (b) report cumulative ablations
normalized to VeOmni, with each bar adding one component of \system{}.
Averaged across the two workloads,
hierarchical token deduplication, replication, replica-gradient synchronization
overlap, communication-aware placement, and the expert-compute term reduce
step time relative to the preceding configuration by $40.3\%$, $12.9\%$, $6.7\%$, $8.5\%$, and $3.6\%$, respectively.
The monotonic improvements show that both the runtime mechanisms and the joint
placement objective contribute to the final result across workloads.

Figure~\ref{fig:ablation-sensitivity}(c) sweeps $B/E$ from $0$ to $1$.
At $B/E=0.25$, throughput is already $1.30\times$ that without additional
copies. Budgets from $0.25$ to $0.75$ provide similar throughput, while peak
allocated memory rises from $32.72$ to $40.12$~GiB. The full 1-fold budget
achieves the best throughput, $1.43\times$, at a peak allocation of
$44.21$~GiB. Thus, $B/E$ exposes a direct performance--memory tradeoff and
can be selected according to the available device memory.

Figure~\ref{fig:ablation-sensitivity}(d) shows that \system{} is
$2.16\times$ faster than VeOmni with a tighter step-time band, indicating
stable load balance as routes evolve. The optimizer runs asynchronously to
refresh $L$ and $M$ every 100 steps, so the 5 updates expose only $58.6$~s in
total, accounting for $0.64\%$ of the 600-step run.

\section{Conclusion}

In this paper, we present \system{}, an end-to-end MoE training system that
jointly optimizes replica allocation, hierarchical expert placement, and
source-aware token-to-copy mapping. Its calibrated objective captures both
duplicate-free hierarchical communication and non-deduplicated expert-compute
load, while affinity-aware hierarchical partitioning and alternating
layout--mapping refinement make the coupled optimization tractable. At runtime,
\system{} asynchronously refreshes the
layout and mapping without restarting training and overlaps replica-gradient
synchronization with attention backward. Evaluations on a 32-GPU cluster and
a 64-NPU cluster using Qwen3-VL and DeepSeek-V3 show that \system{} accelerates
A2A communication by up to $6.94\times$ and end-to-end training by
$1.74$--$2.33\times$ over VeOmni, demonstrating consistent benefits across
hardware platforms and workloads. The component breakdown confirms that
optimizing communication or computation alone leaves complementary
bottlenecks, whereas their joint optimization consistently improves
end-to-end performance. The sensitivity study further exposes a tunable
performance--memory tradeoff, and the 600-step experiment shows that
asynchronous reoptimization preserves the training benefit with limited
exposed update overhead.

\section*{Use of AI Disclosure}
We used Codex (GPT-5.5) for code validation and experimental setups debugging, and we also used Overleaf's default AI tool to help rephrase the writing of sentences. We have fully reviewed and validated all AI-generated suggestions before finalization.

\balance
\bibliographystyle{IEEEtran}
\bibliography{references}

\end{document}